# Completing electron scattering studies with the inert gas column: e - Rn scattering and Ionization

[1]FORAM M JOSHI, [2]K N JOSHIPURA, [3]ASHA S CHAUDHARI, [4]HITESH S. MODI & [5]MANISH J. PINDARIA

[1] G H Patel College of Engineering and Technology, Vallabh Vidyanagar - 388120, INDIA

[2] Sardar Patel University, Vallabh Vidyanagar - 388120, INDIA,

[3] G. D. Higher secondary School, Visnagar - 384315, INDIA

[4] Shree Sardar Patel Higher Secondary School, Patan - 384265, INDIA

[5] Sheth M. N. Science College, Patan - 384265, INDIA

*Key words: Radon atoms, electron impact ionization, CSP-ic method*

## INTRODUCTION

Interest in the inert or noble- gas atoms in general arises because they are ideal as test systems for various theoretical models of electron scattering and also since their interaction processes serve as reference for the determination of instrumental responses in electron scattering experiments. The ionization cross section data of ground state inert gas atoms He through Xe are considered to be benchmark data. Our aim in this paper is to provide theoretical results on electron scattering with Radon atoms, as it would complete the studies on the entire inert gas column. That is possible with this particular column only, in view of the preceding literature on He through Xe . Inert gas radon is radioactive, and would be a difficult target for electron scattering experiments. In the present calculations, the complications arising from radioactivity are not considered. We provide hitherto unavailable cross sections on atomic radon, and also provide opportunity of the comparison of electron impact cross sections over all the inert gas targets.

**THEORETICAL METHODOLOGY**

Let us denote the total (complete) cross section of electron-atom collisions is by $Q_T$, which shows the sum of total elastic cross section $Q_{el}$ and total inelastic cross section $Q_{inel}$. Thus

$$Q_T(E_i) = Q_{el}(E_i) + Q_{inel}(E_i) \tag{1}$$

Further,

$$Q_{inel}(E_i) = \Sigma Q_{ion}(E_i) + \Sigma Q_{exc}(E_i) \tag{2}$$

Where $E_i$ is the incident electron energy. The quantity $\Sigma Q_{ion}(E_i)$ in equation (2) shows the sum-total of first, second etc ionization cross sections of the target. For simplicity we denote the first term simply by $Q_{ion}$. The quantity $\Sigma Q_{exc}(E_i)$ shows the summed total electronic excitation cross sections.

In our publications **[1-6]** on electron–atom/molecule scattering, theoretical efforts have been directed toward extracting the ionization contribution from the total inelastic cross section derived from a complex scattering potential. Presently we have employed the well-established Complex Scattering Potential ionization contribution (CSP-ic) formalism developed in the recent years [**1-6**] to obtain $Q_{ion}$ along with other total cross sections for these targets, at energies $E_i$ from the first ionization threshold to 2 keV. With this background let us outline how the total cross sections $Q_{ion}$ of electron scattering from atomic targets are deduced from $Q_{inel}$ within a broad frame-work of complex potential formalism. In the present range of electron energy, many scattering channels that lead to discrete as well as continuum transitions in the target are open.

We have modified the original absorption model, by considering the threshold energy parameter $\Delta$ of the absorption potential $V_{abs}$ as a slowly varying function of $E_i$ around $I$ as discussed in [1-6]. Briefly, a preliminary calculation is done with a fixed value $\Delta = I$, but the variable $\Delta$ accounts for the screening of the absorption potential in the target charge-cloud region and this has been successful in a number of previous studies. Next, we set up the Schrödinger equation with our modified $V_{abs}$, and find the complex phase shifts $\delta_l = Re\ \delta_l + i\ Im\ \delta_l$ for various partial waves $l$ by following the Variable Phase Approach [7]. The total elastic ($Q_{el}$), inelastic ($Q_{inel}$) and total (complete) cross sections ($Q_T$) are generated from the S- matrix as per the standard expressions [8].

Now, electron impact ionization corresponds to infinitely many open channels, as against the electronic excitation, which comes from a small number of discrete scattering channels. Therefore, the onization channel becomes dominating gradually as the incident energy exceeds $I$, thereby making $Q_{ion}$ the main contribution to $Q_{inel}$. Thus from equation (2), we have in general

$$Q_{inel}(E_i) \geq Q_{ion}(E_i) \qquad (3)$$

There is no rigorous way to project out $Q_{ion}$ from $Q_{inel}$. But in order to determine $Q_{ion}$ from $Q_{inel}$, a reasonable approximation has been evolved by starting with a ratio function,

$$R(E_i) = \frac{Q_{ion}(E_i)}{Q_{inel}(E_i)} \qquad (4)$$

Perhaps a first ever estimate of ionization in relation to excitation processes was made, for water molecules, by Turner et al [9].

The usual complex potential calculations include ionization contribution within the inelastic cross section. In order to deduce the said contribution, we have introduced a method based on the equation **(4).** In our Complex Scattering Potential – ionization contribution (*CSP-ic*) method, the energy dependence of R ($E_i$) is represented by the following relation [1-6].

$$R(E_i) = 1 - C_1\left[\frac{C_2}{U+a} + \frac{\ln(U)}{U}\right] \tag{5}$$

where the incident energy is scaled to the ionization energy $I$ through a dimensionless variable,

$$U = \frac{E_i}{I} \tag{6}$$

Equation (5) involves dimensionless parameters $C_1$, $C_2$, and $a$, which are determined by imposing three conditions on the function $R(E_i)$ as discussed in our papers [1-6]. Briefly, we have $R = 0$ at the ionization threshold and the ratio takes up asymptotic value $R' \approx 1$ at high energies typically above 1000 eV, in view of equation (5). The third condition on $R$ arises from its behaviour at the peak of $Q_{inel}$, and is expressed in the following manner.

$$R(E_i) = \begin{cases} 0, \text{ at } E_i = I \\ R_p, \text{ at } E_i = E_p \\ R', \text{ for } E_i \gg E_p \end{cases} \tag{7}$$

Here, $E_p$ stands for the incident energy at which our calculated inelastic cross section $Q_{inel}$ attains its maximum, while $R_p \cong 0.7$ stands for the value of the ratio $R$ at $E_i = E_p$. The choice of this value is approximate but physically justified. The peak position $E_p$ occurs at an incident energy where the dominant discrete excitation cross sections are on the wane, while the ionization cross section is rising fast, suggesting that the $R_p$ value should be above 0.5 but still below 1. This behavior is attributed to the faster fall of the first term $\sum Q_{exc}$ in equation (2). An exact theoretical evaluation of $R_p$ does not seem to be possible, but one can try to see the effect of a small change in this value. The choice of $R_p$ in equation (7) is not rigorous and it introduces uncertainty in the final results. From equation (6) at high energies, the ratio R' approaches to unity which is physically supported by the low ionization cross sections in the same energy region. We employ the three conditions on R to evaluate the three parameters of equation (5) and take $R_P = 0.70$ initially. The parameters are determined iteratively. Thus we deduce the $Q_{ion}$ from the calculated $Q_{inel}$ by using equation (**4**). The method of complex potential coupled with ionization contribution to inelastic scattering as explained above offers the determination of different total cross sections $Q_T$. In a variant of the usual CSP-ic method, we start by by taking R'≈ 0.95, and impose the conditions mentioned in equation (7). The alternate calculation procedure provides $R_P = 0.719$ which we employ to calculate the parameters a, $C_1$ and $C_2$ hence to obtain $Q_{ion}$ from $Q_{inel}$ using the equation (4).

All the cross sections are examined here as functions of incident electron energy.

**RESULTS AND DISCUSSION**

The present work is important in view of the energy range in which ionization is taking place along with elastic scattering as well as discrete atomic transitions in Rn. In figure 1 we have shown $Q_{ion}$ and $Q_{inel}$ of atomic radon as functions of electron energy. The upper most curve is $Q_{inel}$, and it exhibits the expected energy dependence.

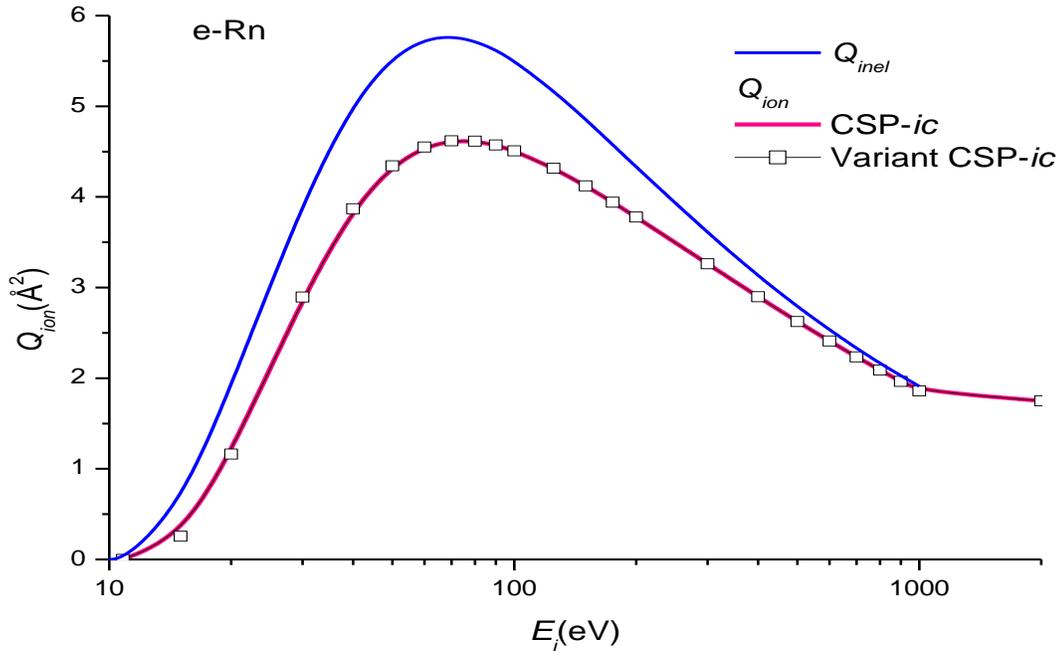

**Figure 1:- ionization cross sections (in Å$^2$) of electron scattering with Radon atoms**

This figure represents a theoretical study not made so far on this atom. With both the methods the ratio Rp at peak of the $Q_{inel}$ is close to 0.70. Hence there is hardly any change in ionization cross sections. Typically above 1000 eV, the $Q_{ion}$ and $Q_{inel}$ are indistinguishable.

We summarize in table 1 an important comparison, in which the various properties and calculated peak cross sections of all the members of inert gas column in periodic table, are displayed. For the inert-gas atoms from He to Xe, theoretical data from [10] are included in this table.

| Target Atom | First ionization threshold eV | Peak position $\varepsilon_{ion}$ eV | Average atomic radius Å | Dipole polarizability Å$^3$ | Peak cross section $\sigma_{max}$ Å$^2$ |
|---|---|---|---|---|---|
| He | 24.6 | 120 | 0.49 | 0.20 | 0.38 |
| Ne | 21.6 | 200 | 0.51 | 0.40 | 0.83 |
| Ar | 15.6 | 100 | 0.87 | 1.64 | 2.54 |
| Kr | 14.0 | 90 | 1.03 | 2.48 | 4.20 |
| Xe | 12.1 | 75 | 1.20 | 4.04 | 5.43 |
| Rn | 10.8 | 65 | 1.34 | 5.30 | 5.77 |

**Table 1: Various properties and calculated cross sections of all the inert gas atoms. Previous cross section data are from [10].**

**CONCLUSIONS**

In conclusion, an interesting study is reported on electron impact ionization of Radon atoms. Complications arising out of radioactive nature of the target are not considered. We have reported theoretical cross sections of electron - Radon collisions for which there are no experimental or theoretical investigations so far. The paper thus presents new results, and seeks to complete electron collision investigations with the inert-gas column of the periodic table.